\documentclass[reprint,superscriptaddress,amsmath,amssymb,aps,prc]{revtex4-1}

\usepackage{graphicx}
\usepackage{hyperref}
\usepackage{siunitx}
\newcommand{\SIhyp}[2]{\SI[number-unit-product={\text{-}}]{#1}{#2}}
\DeclareSIUnit\clight{\text{\ensuremath{c}}}

\begin{document}

\title{Structure of $^{30}$Mg explored via in-beam $\gamma$-ray spectroscopy}

\author{N.~Kitamura}
\email{kitamura@cns.s.u-tokyo.ac.jp}
\affiliation{Center for Nuclear Study, University of Tokyo, Hirosawa 2-1, Wako, Saitama 351-0198, Japan}
\author{K.~Wimmer}
\affiliation{University of Tokyo, Hongo 7-3-1, Bunkyo, Tokyo 113-0033, Japan}
\affiliation{Department of Physics, Central Michigan University, Mt.\ Pleasant, Michigan 48859, USA}
\affiliation{National Superconducting Cyclotron Laboratory, Michigan State University, East Lansing, Michigan 48824, USA}
\altaffiliation{Present address: Instituto de Estructura de la Materia, CSIC, E-28006 Madrid, Spain}
\author{N.~Shimizu}
\affiliation{Center for Nuclear Study, University of Tokyo, Hirosawa 2-1, Wako, Saitama 351-0198, Japan}
\author{V.~M.~Bader}
\affiliation{National Superconducting Cyclotron Laboratory, Michigan State University, East Lansing, Michigan 48824, USA}
\affiliation{Department of Physics and Astronomy, Michigan State University, East Lansing, Michigan 48824, USA}
\author{C.~Bancroft}
\affiliation{Department of Physics, Central Michigan University, Mt.\ Pleasant, Michigan 48859, USA}
\author{D.~Barofsky}
\affiliation{Department of Physics, Central Michigan University, Mt.\ Pleasant, Michigan 48859, USA}
\author{T.~Baugher}
\affiliation{National Superconducting Cyclotron Laboratory, Michigan State University, East Lansing, Michigan 48824, USA}
\affiliation{Department of Physics and Astronomy, Michigan State University, East Lansing, Michigan 48824, USA}
\author{D.~Bazin}
\affiliation{National Superconducting Cyclotron Laboratory, Michigan State University, East Lansing, Michigan 48824, USA}
\author{J.~S.~Berryman}
\affiliation{National Superconducting Cyclotron Laboratory, Michigan State University, East Lansing, Michigan 48824, USA}
\author{V.~Bildstein}
\affiliation{Department of Physics, University of Guelph, Guelph, Ontario N1G 2W1, Canada}
\author{A.~Gade}
\affiliation{National Superconducting Cyclotron Laboratory, Michigan State University, East Lansing, Michigan 48824, USA}
\affiliation{Department of Physics and Astronomy, Michigan State University, East Lansing, Michigan 48824, USA}
\author{N.~Imai}
\affiliation{Center for Nuclear Study, University of Tokyo, Hirosawa 2-1, Wako, Saitama 351-0198, Japan}
\author{T.~Kr\"oll}
\affiliation{Institut f\"ur Kernphysik, Technische Universit\"at Darmstadt, 64289 Darmstadt, Germany}
\author{C.~Langer}
\affiliation{National Superconducting Cyclotron Laboratory, Michigan State University, East Lansing, Michigan 48824, USA}
\author{J.~Lloyd}
\affiliation{Department of Physics, Central Michigan University, Mt.\ Pleasant, Michigan 48859, USA}
\author{E.~Lunderberg}
\affiliation{National Superconducting Cyclotron Laboratory, Michigan State University, East Lansing, Michigan 48824, USA}
\affiliation{Department of Physics and Astronomy, Michigan State University, East Lansing, Michigan 48824, USA}
\author{G.~Perdikakis}
\affiliation{Department of Physics, Central Michigan University, Mt.\ Pleasant, Michigan 48859, USA}
\affiliation{National Superconducting Cyclotron Laboratory, Michigan State University, East Lansing, Michigan 48824, USA}
\author{F.~Recchia}
\affiliation{National Superconducting Cyclotron Laboratory, Michigan State University, East Lansing, Michigan 48824, USA}
\author{T.~Redpath}
\affiliation{Department of Physics, Central Michigan University, Mt.\ Pleasant, Michigan 48859, USA}
\author{S.~Saenz}
\affiliation{Department of Physics, Central Michigan University, Mt.\ Pleasant, Michigan 48859, USA}
\author{D.~Smalley}
\affiliation{National Superconducting Cyclotron Laboratory, Michigan State University, East Lansing, Michigan 48824, USA}
\author{S.~R.~Stroberg}
\affiliation{National Superconducting Cyclotron Laboratory, Michigan State University, East Lansing, Michigan 48824, USA}
\affiliation{Department of Physics and Astronomy, Michigan State University, East Lansing, Michigan 48824, USA}
\author{J.~A.~Tostevin}
\affiliation{Department of Physics, University of Surrey, Guildford, Surrey GU2 7XH, United Kingdom}
\author{N.~Tsunoda}
\affiliation{Center for Nuclear Study, University of Tokyo, Hirosawa 2-1, Wako, Saitama 351-0198, Japan}
\author{Y.~Utsuno}
\affiliation{Advanced Science Research Center, Japan Atomic Energy Agency, Tokai, Ibaraki 319-1195, Japan}
\affiliation{Center for Nuclear Study, University of Tokyo, Hirosawa 2-1, Wako, Saitama 351-0198, Japan}
\author{D.~Weisshaar}
\affiliation{National Superconducting Cyclotron Laboratory, Michigan State University, East Lansing, Michigan 48824, USA}
\author{A.~Westerberg}
\affiliation{Department of Physics, Central Michigan University, Mt.\ Pleasant, Michigan 48859, USA}

\date{\today}

\begin{abstract}
\begin{description}
\item[Background] In the ``island of inversion'', ground states of neutron-rich $sd$-shell nuclei exhibit strong admixtures of intruder configurations from the $fp$ shell. The nucleus $^{30}$Mg, located at the boundary of the island of inversion, serves as a cornerstone to track the structural evolution as one approaches this region.
\item[Purpose] Spin-parity assignments for excited states in $^{30}$Mg, especially negative-parity levels, have yet to be established. In the present work, the nuclear structure of $^{30}$Mg was investigated by in-beam $\gamma$-ray spectroscopy mainly focusing on firm spin-parity determinations.
\item[Method] High-intensity rare-isotope beams of $^{31}$Mg, $^{32}$Mg, $^{34}$Si, and $^{35}$P bombarded a Be target to induce nucleon removal reactions populating states in $^{30}$Mg. $\gamma$ rays were detected by the state-of-the-art $\gamma$-ray tracking array GRETINA.
For the direct one-neutron removal reaction, final-state exclusive cross sections and parallel momentum distributions were deduced. Multi-nucleon removal reactions from different projectiles were exploited to gain complementary information.
\item[Results] With the aid of the parallel momentum distributions, an updated level scheme with revised spin-parity assignments was constructed. Spectroscopic factors associated with each state were also deduced.
\item[Conclusions] Results were confronted with large-scale shell-model calculations using two different effective interactions, showing excellent agreement with the present level scheme. However, a marked difference in the spectroscopic factors indicates that the full delineation of the transition into the island of inversion remains a challenge for theoretical models.
\end{description}
\end{abstract}


\maketitle

\section{Introduction}

Changes from the canonical shell structure have been observed in neutron-rich unstable nuclei~\cite{SOR08}. The unexpected deformation of the ground states of unstable Ne, Na, and Mg isotopes located around the classical magic number $N=20$ has attracted much attention since the 1970s. Triggered by the pioneering experiment that revealed the unexpectedly high binding energies in Na isotopes~\cite{THI75}, anomalies were also found in the binding energies of Ne and Mg isotopes~\cite{ORR91,DET83}. The low excitation energies of the first excited $2^+$ states~\cite{DET79,IWA01,YAN03} and the large collectivity in $^{32}$Mg~\cite{MOT95} have added experimental evidence for the disappearance of the $N=20$ magic number. In the shell-model picture, the disappearance of magicity in this region is driven by the reduction of the effective single-particle energy gap between the $sd$ and $pf$ orbitals corresponding to the $N=20$ gap~\cite{SOR08,OTS05,OTS10}. This leads to increased contributions of excitations across the gap, e.g.\ two-particle two-hole (2p2h) configurations, in the ground states of these nuclei. The region where these unusual intruder configurations are dominant in the ground states was named the ``island of inversion''~\cite{WAR90}. This region has provided a rich testing ground for the modeling of nuclear structure far away from stability and particularly the evolution of the shell structure as a function of proton and neutron number.

The full delineation of the reasons for the emergence of the island of inversion is still underway. For the Mg isotopes, the characteristics of the ground state of $^{31}$Mg were studied by hyperfine structure and $\beta$-NMR measurements. It was found that the ground state has an unexpected spin-parity of $1/2^+$, in contrast to the na\"ive expectation of $3/2^+$ based on the normal configuration, i.e.\ a single neutron hole in the $1d_{3/2}$ orbital. Accordingly, the ground state is understood as having a particle-hole dominated intruder configuration, thus placing $^{31}$Mg on the border of the island of inversion~\cite{NEY05}. To explore the onset of intruder configurations, one-neutron knockout measurements on $^{30}$Mg and $^{32}$Mg, respectively, leading to $^{29}$Mg and $^{31}$Mg, were performed~\cite{TER08}. The measurements revealed that there is a significant increase in the $fp$-shell occupation at $^{32}$Mg compared to $^{30}$Mg, pointing to a substantial structural change with the addition of two neutrons. Another important facet of the transition into the island of inversion is shape coexistence, where different shapes appear in near-degenerate states in the same nucleus. Experimental indications for shape coexistence have been found in $^{30}$Mg~\cite{SCH09} and $^{32}$Mg~\cite{WIM10,ELD19}. The ground state of $^{30}$Mg is characterized as having an essentially normal, spherical configuration where all the active neutrons are confined to the $sd$ shell (0p0h), while the excited $0^+$ state is described as an intruder, deformed configuration where two neutrons are promoted across the $N = 20$ gap (2p2h). The situation is considered to be inverted for $^{32}$Mg, but recent theoretical and experimental studies~\cite{CAU14,TSU17,MAC16,ELD19} suggested that the above-mentioned interpretation might be too simplified. In this context, it is important to establish a microscopic view of $^{30}$Mg.

To date, much experimental effort has been devoted to investigating the structure of $^{30}$Mg. Since the first observations of $^{30}$Mg~\cite{ART71,ROE74}, excited levels in this nucleus have been investigated by $\beta$-$\gamma$ measurements of $^{30}$Na~\cite{DET79,GUI84,BAU89,SHI14}. Levels in $^{30}$Mg were also studied by $\beta$-delayed one-neutron emission from $^{31}$Na and two-neutron emission from $^{32}$Na~\cite{KLO93,MAC05,MAT07}. From these studies, the $2_1^+$ state was established at \SI{1482}{keV} and the shape coexisting $0_2^+$ state was identified at \SI{1789}{keV}~\cite{SCH09}. The measured monopole transition strength of $\rho^2(E0; 0_2^+ \to 0_1^+) = \num[separate-uncertainty]{26.2 +- 7.5 e-3}$ was moderate and interpreted as arising from a small mixing of configurations with very different deformation~\cite{SCH09}. To investigate the ground-state collectivity, three Coulomb excitation measurements on $^{30}$Mg have been performed~\cite{PRI00,CHI01,NIE05}. The latest measurement employing ``safe'' Coulomb excitation gave a moderate collectivity of $B(E2;0_1^+ \to 2_1^+) = \SI{241(31)}{\elementarycharge^2.fm^4}$. Recently, in-beam $\gamma$-ray measurements on $^{30}$Mg using the $^{14}\text{C}(^{18}\text{O},2p)^{30}\text{Mg}$ fusion-evaporation reaction~\cite{DEA10} and the one-neutron knockout reaction from $^{31}$Mg~\cite{FER18} have been reported. These measurements proposed candidates for negative-parity states lying at unusually low excitation energy around \SI{2.5}{MeV}. The negative-parity states are formed by promoting an odd number of neutrons from the $sd$ shell to the higher-lying $fp$ orbitals with opposite parity. Thus, the location of negative-parity states is particularly important to discuss the structural evolution approaching the island of inversion~\cite{SOR08,IWA00}, as it is indicative of the effective size of the $N=20$ gap. It should also be emphasized that these negative-parity candidates are lying at lower energies than those presently known in $^{32}$Mg~\cite{TRI08}, located at the heart of the island of inversion, implying an unusual structural change taking place at $^{30}$Mg. Moreover, these energies are at variance with shell-model calculations~\cite{DEA10,FER18}, posing a question concerning the available shell-model description of $^{30}$Mg. A more detailed spectroscopic study for conclusive spin-parity assignments was thus demanded.

In the present work, the one-neutron knockout reaction from $^{31}$Mg induced by a nuclear target was performed to enable firm spin-parity assignments to states in $^{30}$Mg. Additionally, the two-neutron removal reaction from $^{32}$Mg was employed to populate states in $^{30}$Mg. This reaction will receive only a small contribution from direct knockout, and therefore a near-statistical population of states with some intruder contributions is expected. Multi-nucleon removal reactions starting from $^{34}$Si and $^{35}$P, both of which lie outside the island of inversion, are also exploited. In these fragmentation-like reactions, the level population is less sensitive to the nuclear structure, but high-spin states, possibly with the yrast nature, are known to be preferentially populated~\cite{DEJ97,BEL00,YON01,CRA16}. Along with the spectroscopic factors that can be deduced from the one-neutron knockout measurement, a new level scheme with updated spin-parity assignments serves as an important test to the shell model and furthers our understanding.

\section{Experimental details}

\begin{figure*}[tb]
\includegraphics[scale=0.5]{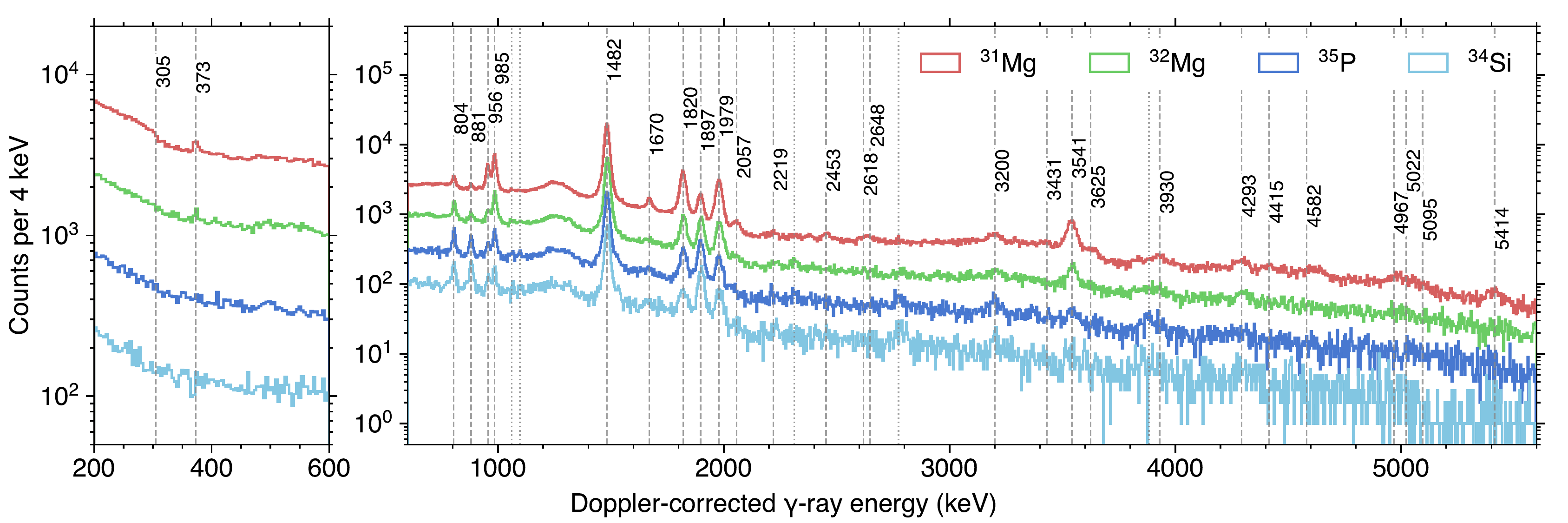}
\caption{Doppler-corrected add-back $\gamma$-ray spectra of $^{30}$Mg taken from four different reaction channels. The red, green, blue, and cyan histograms respectively show the spectra obtained from the incoming projectiles of $^{31}$Mg, $^{32}$Mg, $^{35}$P, and $^{34}$Si. Transitions observed in the one-neutron removal reaction from $^{31}$Mg are indicated by dashed lines and labeled with their energies. Transitions with energies of \num{1061(3)}, \num{1097(2)}, \num{2312(4)}, \num{2774(5)}, and \SI{3883(9)}{keV} are observed in multi-nucleon removal reactions and indicated by dotted lines.}
\label{fig:gamma}
\end{figure*}

The rare-isotope beams were produced by fragmentation of a $^{48}$Ca primary beam at \SI{140}{MeV/nucleon} delivered by the Coupled Cyclotron Facility at the National Superconducting Cyclotron Laboratory at Michigan State University~\cite{GAD16}. The primary beam impinged on a \SIhyp{846}{mg/cm^2} thick $^9$Be production target, and the rare-isotope ions of interest were separated from other fragments and selectively transmitted through the A1900 fragment separator~\cite{MOR03} with a momentum acceptance of \SI{1.1}{\percent}. An aluminum wedge degrader with a thickness of \SI{300}{mg/cm^2} was placed at the intermediate plane of the A1900 to purify the cocktail of the fragments. The secondary beams containing $^{31}$Mg, $^{32}$Mg, $^{34}$Si, and $^{35}$P were directed onto a \SIhyp{375}{mg/cm^2} thick $^9$Be reaction target to induce nucleon removal reactions to populate states in $^{30}$Mg. The identification of incoming radioactive ions was accomplished by measuring the time-of-flight (TOF) between two plastic scintillators placed at the extended focal plane of the A1900 and the object position of the analysis beam line of the S800 spectrograph located just upstream of the reaction target. The beam energies, intensities, and purities are tabulated in Table~\ref{tab:beams}.

\begin{table}[!tb]
\centering
\caption{List of rare-isotope beams produced in the measurement. Their incident energies (\si{MeV/nucleon}), averaged intensities on target (\si{s^{-1}}), and purities (\si{\percent}) are shown together.}
\label{tab:beams}
\begin{tabular}{lcccc}
\toprule
Secondary beam      & $^{31}$Mg   & $^{32}$Mg   & $^{34}$Si   & $^{35}$P    \\
\colrule
Energy              & \num{97.9}  & \num{99.1}  & \num{94.8}  & \num{102.3} \\
Intensity           & \num{2.1e4} & \num{6.1e3} & \num{5.2e5} & \num{2.3e5} \\
Purity              & \num{28}    & \num{34}    & \num{66}    & \num{29}    \\
\botrule
\end{tabular}
\end{table}

The reaction products, including $^{30}$Mg, were transmitted to the S800 spectrograph~\cite{BAZ03} and identified by the TOF-$\Delta E$ method. The TOF was measured between two plastic scintillators located at the object position and the focal plane of the S800 spectrograph. This scintillator was also used to trigger the electronics and data acquisition in coincidence with the $\gamma$-ray detection. An ionization chamber installed in the S800 focal plane was used to provide the energy loss information ($\Delta E$) for the identification of the atomic number. The trajectories of the reaction products were tracked in the S800 focal plane by two cathode readout drift chambers, providing position and angle information~\cite{YUR99}. The parallel momentum and the angle at the target position were reconstructed from the trajectory measurements in the focal plane by the inverse-map approach~\cite{BER93}.

The reaction target was surrounded by the Gamma-Ray Energy Tracking In-beam Nuclear Array (GRETINA)~\cite{PAS13,WEI17} to detect prompt de-excitation $\gamma$ rays emitted by the reaction products in-flight. The array consisted of the seven modules of GRETINA at the time of the experiment. Each module houses four closely-packed crystals of high-purity Ge. The electrodes on the crystal are segmented into 36 elements so that interaction points of a $\gamma$ ray can be reconstructed on the sub-segment level by online pulse-shape decomposition. Four modules covered forward angles around \SI{58}{\degree}, while the rest were aligned on the \SI{90}{\degree} ring. The $\gamma$-ray interaction points from GRETINA together with the velocity and angle of the reaction product at the target position, reconstructed from the trajectory measurement in the S800, enabled the Doppler correction of $\gamma$ rays on an event-by-event basis. In the present measurement, an energy resolution of \SI{1.4}{\percent} in FWHM was achieved. The leading contributions to the linewidth are the uncertainties in the $\gamma$-ray interaction position and the velocity at which $\gamma$ rays are emitted~\cite{WEI17}. The add-back procedure was used for the identification of peaks and the $\gamma$-$\gamma$ analyses. The extraction of $\gamma$-ray intensities was accomplished without add-back.

\section{Analysis and experimental results}

\subsection{Level scheme}

The large statistics collected in the present experiment allowed us to extend the previously known level scheme of $^{30}$Mg. The Doppler-corrected $\gamma$-ray spectra recorded in coincidence with residual $^{30}$Mg nuclei obtained in the different reaction channels, i.e.\ the one-neutron knockout reaction from $^{31}$Mg and the multi-nucleon removal reactions from $^{32}$Mg, $^{34}$Si, and $^{35}$P, are shown in Fig.~\ref{fig:gamma}. The Doppler correction was performed with the mid-target velocity of the residue. The observed peaks are tabulated in Table \ref{tab:gamint}. The most prominent peak, observed at \SI{1482}{keV}, corresponds to the $2_1^+\to0_1^+$ transition. The shoulder-like structure around \SI{305}{keV}, which appears most prominently in one-neutron knockout, is attributed to the $0_2^+\to2_1^+$ transition. Since the $0_2^+$ has a long half-life of \SI{3.9(4)}{ns}~\cite{MAC05}, the $\gamma$-ray emission takes place significantly downstream of the target and thus the $0_2^+\to2_1^+$ transition does not produce a clear peak in the spectrum. The transition energy and lifetime of this $\gamma$ ray was not determined in this work, as the experiment was not designed for measurements of such a long-lived state. The peaks at \SIlist[list-units = single]{804;956;985;1670;1820;1897;1979;3541}{keV} observed in the one-neutron knockout reaction confirm the results of the previous measurement~\cite{FER18}.

\begin{figure}[tb]
\includegraphics[scale=0.5]{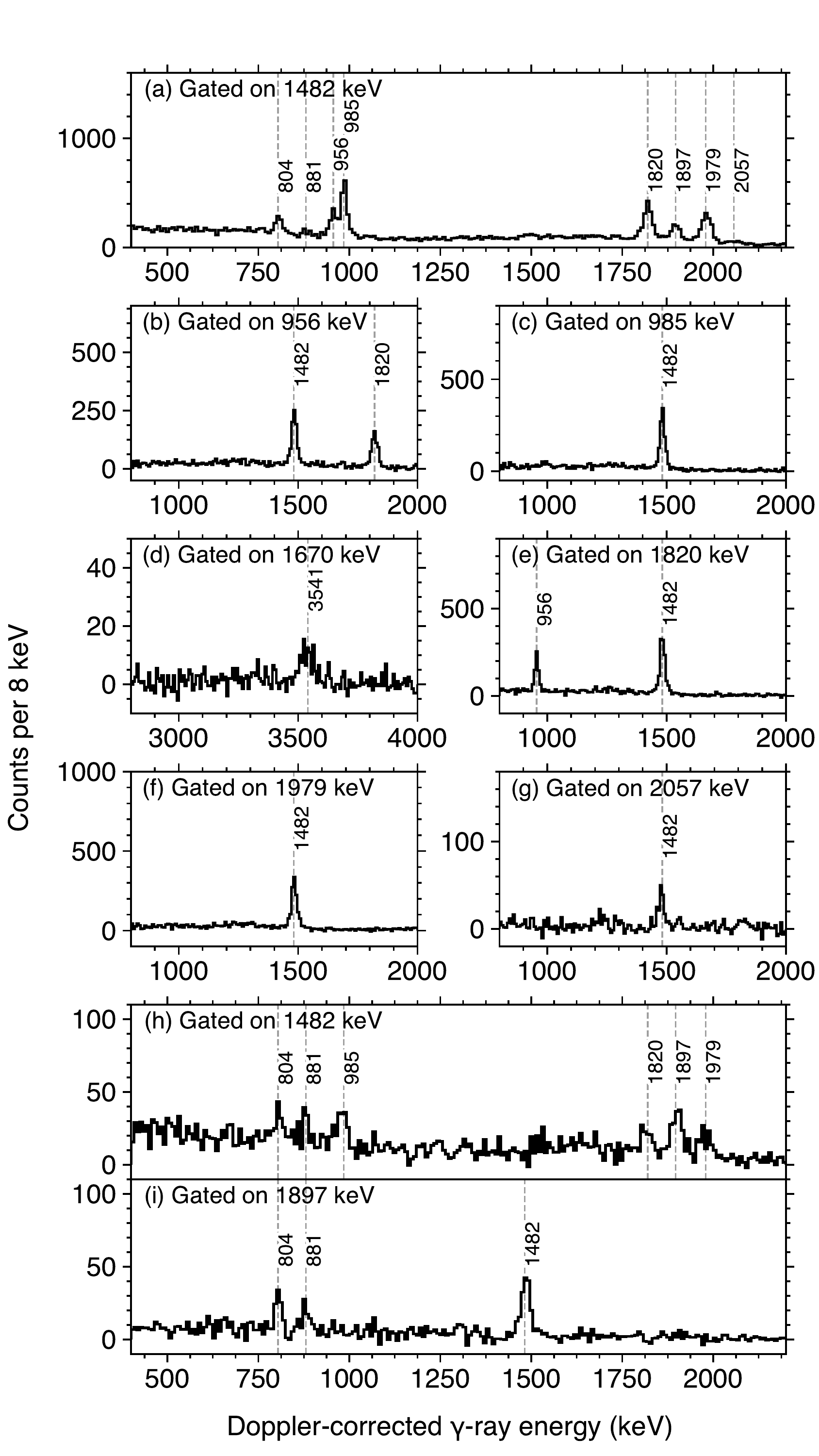}
\caption{Background-subtracted projections of the $\gamma$-$\gamma$ coincidence matrices created by gating on different transitions. Peaks are labeled by their transition energy in \si{keV}. To gain statistics, the $\gamma$-$\gamma$ matrices of the one-neutron knockout reaction from $^{31}$Mg and the two-neutron removal reaction from $^{32}$Mg are added in the panels (a)--(g). The $\gamma$-$\gamma$ matrices of the multi-nucleon removal reactions from $^{34}$Si and $^{35}$P are added in the panels (h) and (i). }
\label{fig:gamgam}
\end{figure}

\begin{figure}[tb]
\includegraphics[scale=0.5]{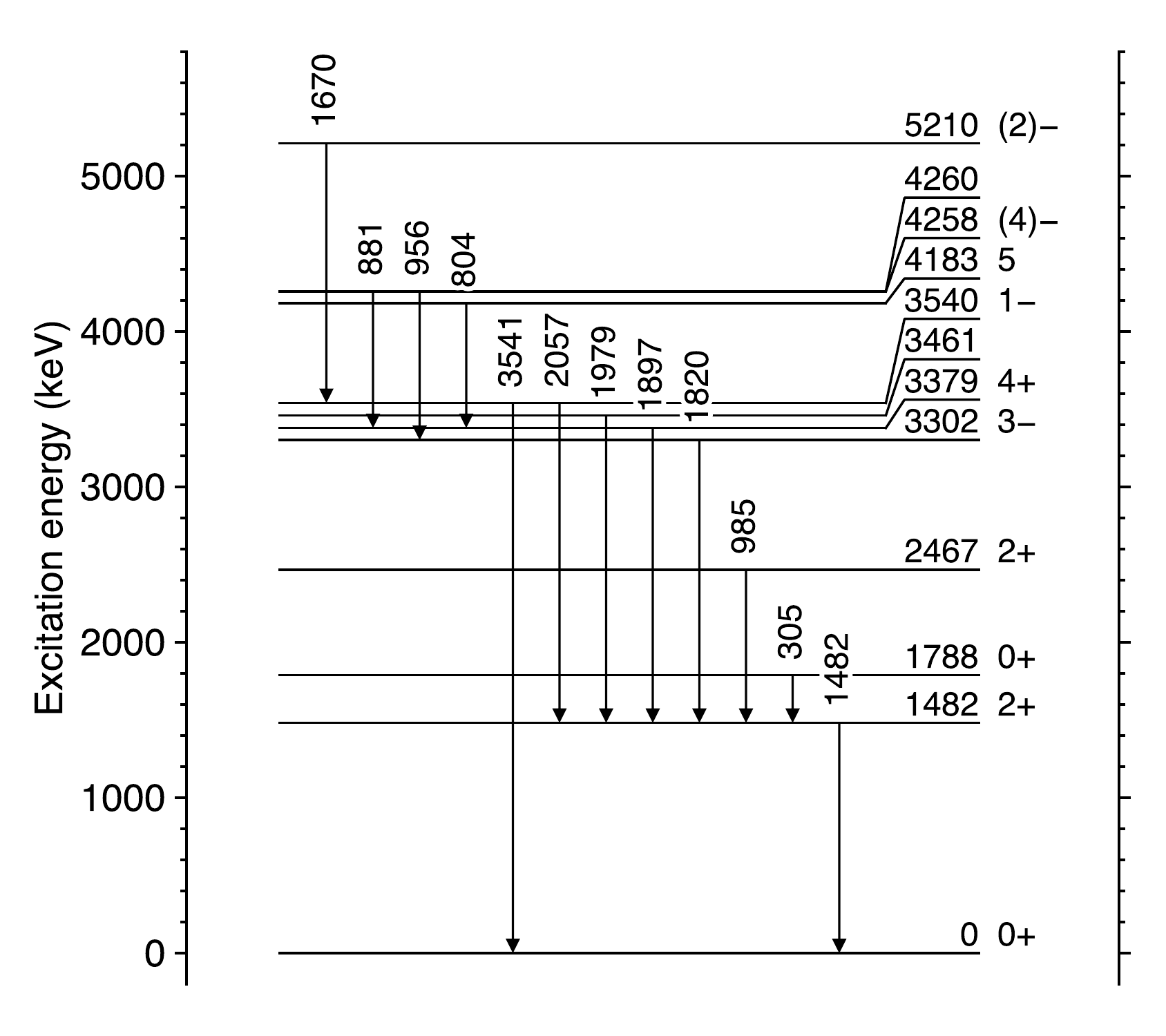}
\caption{Proposed level scheme for $^{30}$Mg from the present work. Excitation energies in keV, spins and parities are shown beside the levels. The spin-parity assignments are discussed in Sec.~\ref{sec:spinparity}. A doublet is proposed at \SI{3461}{keV} (see text for details).}
\label{fig:levels}
\end{figure}

Selected $\gamma$-ray spectra from the $\gamma$-$\gamma$ coincidence analysis are displayed in Fig.~\ref{fig:gamgam}. Based on these observed coincidences, an updated level scheme has been constructed, as shown in Fig.~\ref{fig:levels}. The present level scheme agrees with those constructed in previous studies~\cite{BAU89,SHI14,FER18}. The \SIhyp{1670}{keV} transition, which could not be placed in the level scheme in the previous one-neutron knockout measurement~\cite{FER18}, was found to feed the state at \SI{3540}{keV}, pointing to the existence of a new state at \SI{5210}{keV}. The \SIhyp{2057}{keV} transition is assumed to depopulate the state at \SI{3540}{keV}, guided by $\beta$-$\gamma$ measurements~\cite{BAU89,SHI14}. The \SIhyp{804}{keV} and \SIhyp{881}{keV} transitions feed the state at \SI{3379}{keV}, and these branches are consistent with the level scheme proposed in the previous fusion-evaporation experiment~\cite{DEA10}. Part of the level scheme proposed in Ref.~\cite{DEA10}, especially the $\num{5311}\to\num{4357}\to\num{2541}\to\num{1481}\to\text{g.s.}$ cascade, is at variance with the present results, and the presence of a negative-parity state at \SI{2541}{keV} is not supported. Even though the states at \SI{4260}{keV} and \SI{4258}{keV} lie close to each other in energy, these two states must be different, as the ratio of the \SIhyp{881}{keV} and \SIhyp{956}{keV} $\gamma$-ray intensities, depopulating the two states, strongly depends on the reaction channel. It is also found that the momentum distributions of the reaction residue gated on these two transitions are distinct (see Sec.~\ref{sec:spinparity} for details).

\subsection{Cross sections}

For the one-neutron knockout reaction, the inclusive cross section to all bound states was determined to be \SI{97(3)}{mb}. The target thickness tolerance (\SI{1}{\percent}) and the fluctuation in the secondary beam composition (\SI{3}{\percent}) contribute to the quoted uncertainty. The present result compares well with the previously measured value of \SI{90(12)}{mb}~\cite{FER18}, performed at a lower beam energy with a carbon target.

\begin{figure}[tb]
\includegraphics[scale=0.5]{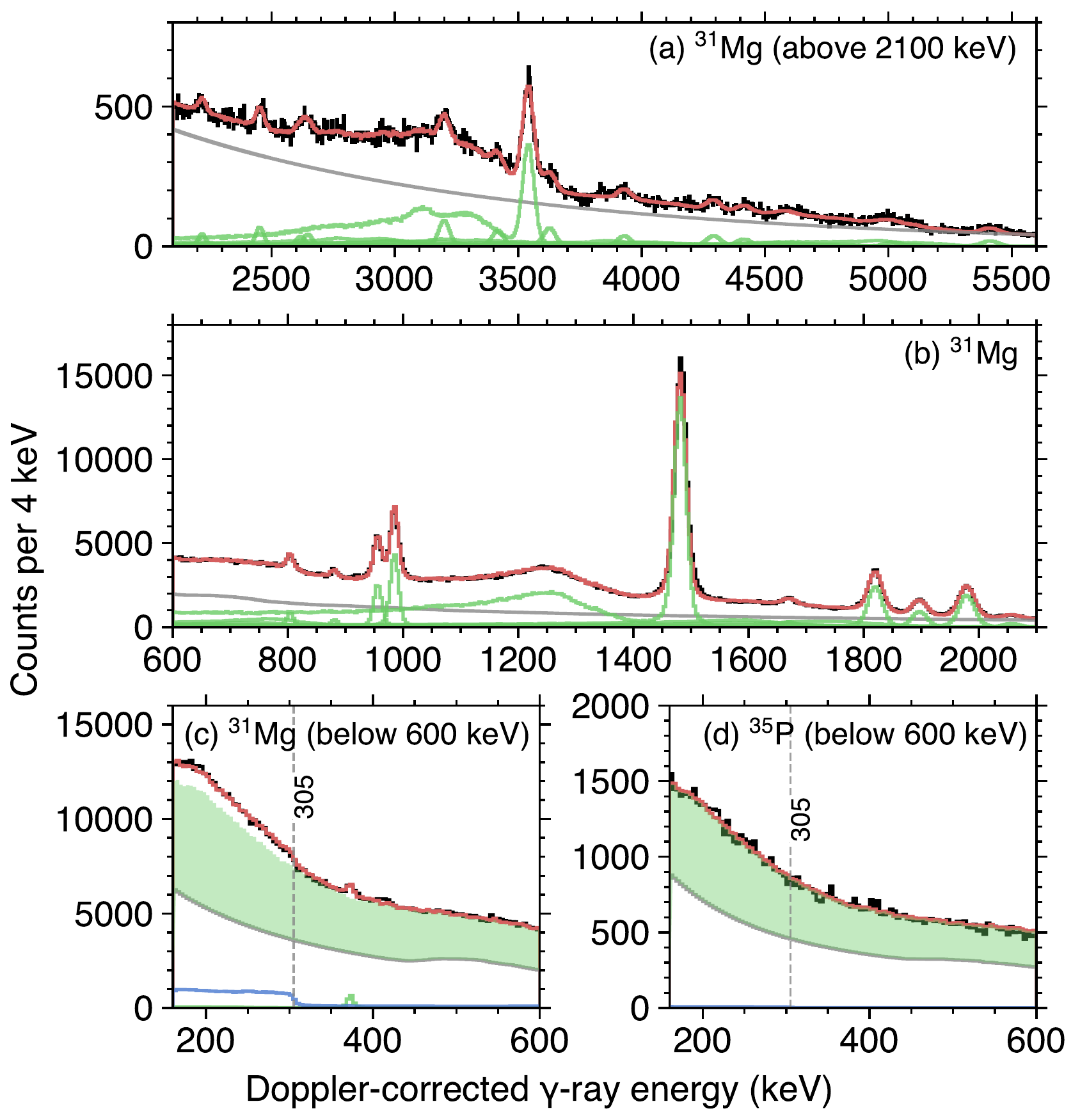}
\caption{Doppler-corrected $\gamma$-ray spectra without add-back (black) for incoming beams of (a--c) $^{31}$Mg and (d) $^{35}$P compared to the best-fit spectra obtained from simulations (red). Individual response functions are indicated by the green histograms. The response function for the delayed \SIhyp{305}{keV} transition is highlighted in blue. The gray solid line denotes background contributions (see text for details). The green shaded area in the panels (c) and (d) represents the sum of the Compton-scattering components of all higher-lying peaks.}
\label{fig:fit}
\end{figure}

In order to deduce the exclusive cross section associated with each state, the observed Doppler-corrected $\gamma$-ray spectrum was fitted by a combination of GRETINA response functions generated by Monte Carlo simulations~\cite{UCGretina}, yielding the $\gamma$-ray intensities (see Fig.~\ref{fig:fit}). In addition to the transitions from the de-excitation of the residual nucleus, a continuous prompt $\gamma$-ray background is commonly observed in in-beam measurements. A double exponential was used to account for this contribution. Discrete $\gamma$-ray peaks observed in the lab-frame spectrum, arising from neutron-induced reactions on Ge crystals, as well as surrounding materials like the Al beam pipe and the Al detector housing, were Doppler-shifted and added to the fit function. The $\gamma$-ray intensities of the peaks lying above \SI{600}{keV} were obtained from a single fit to the spectrum. The low-energy part of the spectrum, around the delayed \SIhyp{305}{keV} transition, was fitted separately with the fixed Compton-scattering components of the higher-lying peaks above \SI{600}{keV} to better model the background contribution. The \SIhyp{305}{keV} response function was generated with a fixed half-life of \SI{3.9}{ns}. The resulting $\gamma$-ray intensities, tabulated in Table~\ref{tab:gamint}, were then corrected for the feeding based on the level scheme constructed in the present work. Some of the transitions could not be placed in the level scheme in the present $\gamma$-$\gamma$ analysis, but in some cases the level scheme constructed in a previous $\beta$-decay measurement~\cite{SHI14} could be used to determine the feeding contributions. The exclusive cross sections are presented in Table~\ref{tab:sfactors} and Fig.~\ref{fig:xsec}. The ground-state cross section was obtained by subtracting the cross section to all excited states from the inclusive cross section. The unplaced transitions listed in Table~\ref{tab:gamint}, with the summed intensity corresponding to a cross section of \SI{2.5(2)}{mb}, were not taken into account when calculating the exclusive cross sections.

\begin{table}[tb]
\centering
\caption{Observed $\gamma$-ray energies for the one-neutron knockout reaction from $^{31}$Mg and their placements in $^{30}$Mg. Intensities relative to the prompt component of the \SIhyp{1482}{keV} transition in the one-neutron knockout spectrum are also indicated. The energies are shown in units of \si{keV}. In cases where the present statistics does not allow for $\gamma$-$\gamma$ coincidences, the placements of transitions into the level scheme are adopted from Ref.~\cite{SHI14} and shown in the parentheses (see text for details). The quoted uncertainties of the relative intensities show the statistical contributions.}
\label{tab:gamint}
\begin{tabular}{S[table-align-text-post=false]cS}
\toprule
{Transition energy\footnotemark[1]} & Placement & {Relative intensity} \\
\colrule
305                        & 1788$\to$1482   & 0.268(17){\footnotemark[2]} \\
373(1)  {\footnotemark[3]} &                 & 0.009( 1) \\
804(1)                     & 4183$\to$3379   & 0.028( 1) \\
881(1)                     & 4260$\to$3379   & 0.016( 1) \\
956(1)                     & 4258$\to$3302   & 0.102( 1) \\
985(1)                     & 2467$\to$1482   & 0.180( 2) \\
1482(2)                    & 1482$\to$0      & 1.000( 2){\footnotemark[4]} \\
1670(3)                    & 5210$\to$3540   & 0.027( 1) \\
1820(3)                    & 3302$\to$1482   & 0.243( 2) \\
1897(3)                    & 3379$\to$1482   & 0.099( 2) \\
1979(3)                    & 3461$\to$1482   & 0.215( 2) \\
2057(3)                    & 3540$\to$1482   & 0.026( 1) \\
2219(5)                    & (4683$\to$2466) & 0.006( 1) \\
2453(4) {\footnotemark[3]} &                 & 0.013( 1) \\
2618(12)                   & (5095$\to$2466) & 0.007( 2) \\
2648(11){\footnotemark[3]} &                 & 0.009( 2) \\
3200(5)                    & (4683$\to$1482) & 0.027( 2) \\
3431                       & (5897$\to$2466) & 0.019( 2) \\
3541(5)                    & 3540$\to$0      & 0.127( 2) \\
3625                       & (5414$\to$1788) & 0.024( 2) \\
3930                       & (5414$\to$1482) & 0.016( 2) \\
4293(7) {\footnotemark[3]} &                 & 0.019( 1) \\
4415                       & (5897$\to$1482) & 0.014( 2) \\
4582                       & (6064$\to$1482) & 0.004( 2) \\
4967                       & (4967$\to$0)    & 0.004( 2) \\
5022                       & (5022$\to$0)    & 0.014( 2) \\
5095                       & (5095$\to$0)    & 0.008( 2) \\
5414                       & (5414$\to$0)    & 0.027( 2) \\
\botrule
\end{tabular}
\footnotetext[1]{Values without uncertainties are taken from the $\beta$-decay study of Ref.~\cite{SHI14}.}
\footnotetext[2]{An additional systematic uncertainty of \SI{10}{\percent} arising from the half-life determination (\SI{3.9(4)}{ns}~\cite{MAC05}) also contributes.}
\footnotetext[3]{Unplaced transitions.}
\footnotetext[4]{The intensity of the \SIhyp{1482}{keV} transition includes only the prompt component.}
\end{table}

\begin{figure}[tb]
\includegraphics[scale=0.5]{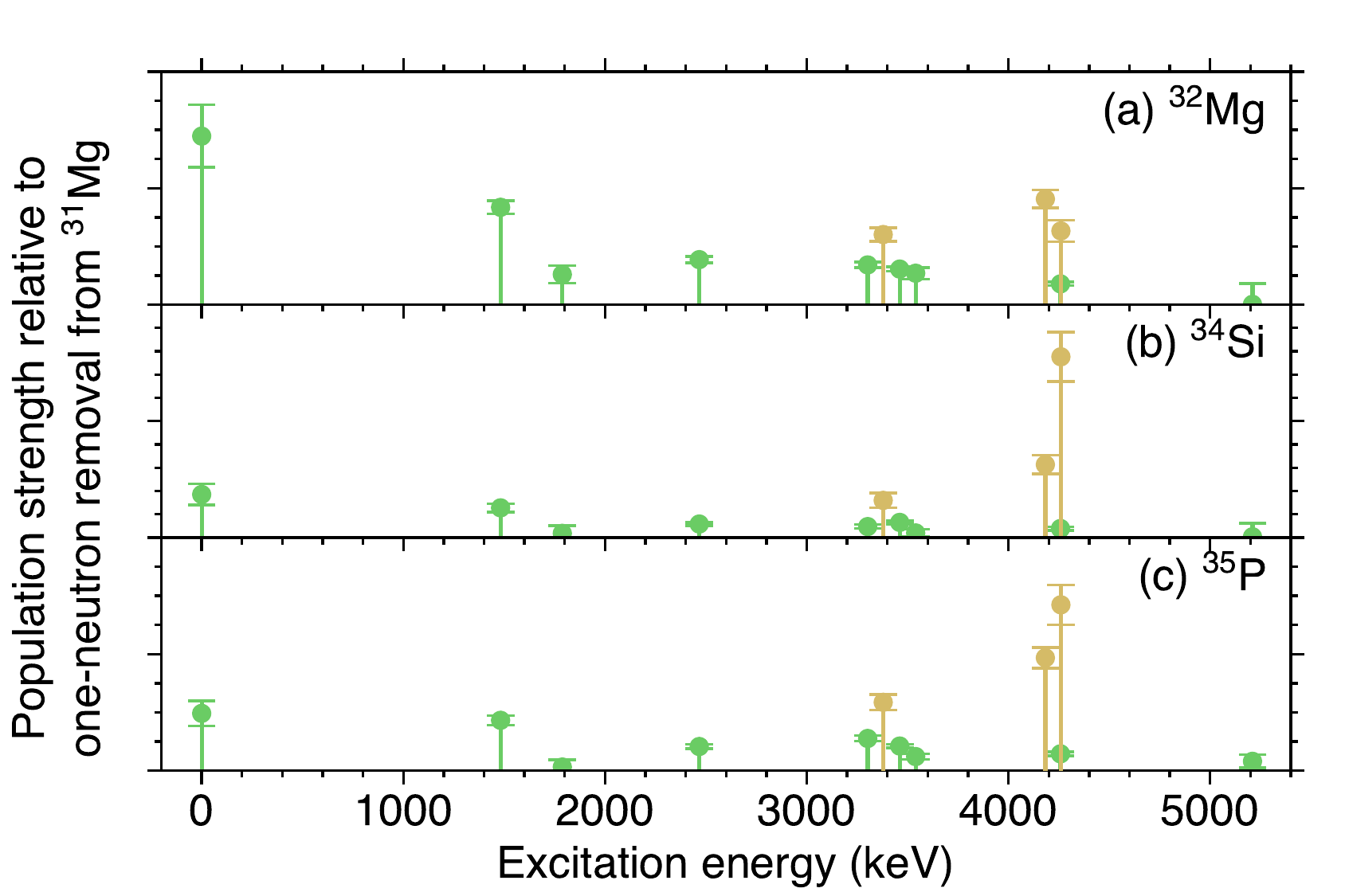}
\caption{Population strengths in the multi-nucleon removal reactions from (a) $^{32}$Mg, (b) $^{34}$Si, and (c) $^{35}$P relative to those of $^{31}$Mg (see Fig.~\ref{fig:xsec} for the exclusive cross sections). Candidates for the high-spin states at \SI{3379}{keV}, \SI{4183}{keV}, and \SI{4260}{keV} are highlighted in yellow.}
\label{fig:relxsec}
\end{figure}

The inclusive cross section for the two-neutron removal reaction from $^{32}$Mg was determined to be \SI{108(2)}{mb}, a larger value than in one-neutron knockout. Similar trends are observed, for example, in Ref.~\cite{VAQ19}. The absolute cross sections for the multi-nucleon removal reactions from $^{34}$Si and $^{35}$P could not be deduced, because the residual $^{30}$Mg nuclei were at the edge of the acceptance of the S800. The $\gamma$-ray spectra for the two-neutron and multi-nucleon removal reactions were analyzed in the same manner as the spectrum obtained for the knockout reaction from $^{31}$Mg. The relative population strengths are shown in Fig.~\ref{fig:relxsec}.

\subsection{Spin-parity assignments}
\label{sec:spinparity}

\begin{figure}[tb]
\includegraphics[scale=0.5]{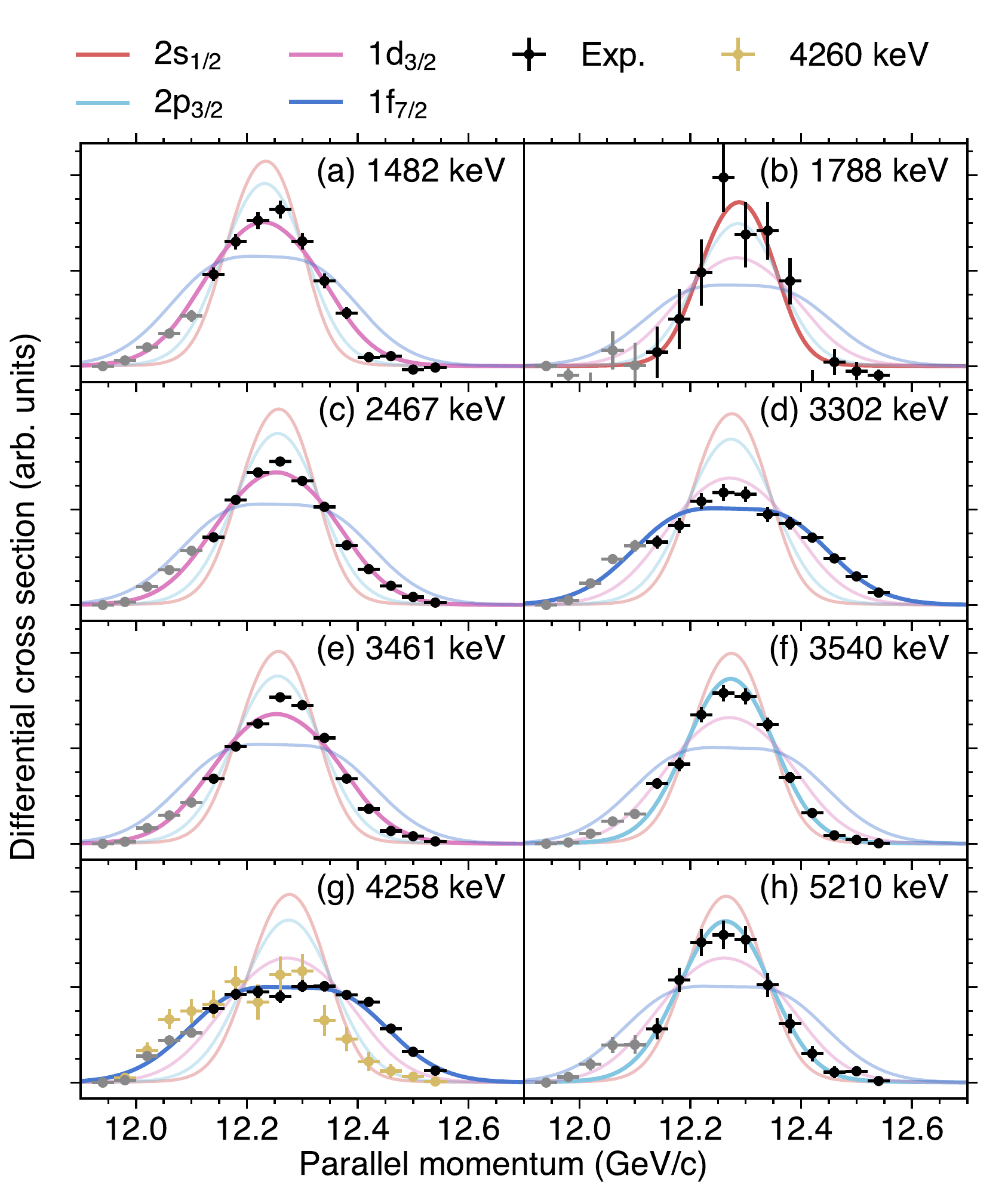}
\caption{Parallel momentum distributions extracted from the experimental data compared with the reaction model calculations. The calculations assume knockout from the $2s_{1/2}$ (red), $1d_{3/2}$ (pink), $1f_{7/2}$ (blue), and $2p_{3/2}$ (cyan) orbitals. The yellow data points in the panel (g) represent the \SIhyp{4260}{keV} momentum distribution. See text for details.}
\label{fig:momdist}
\end{figure}

The spin-parity assignments are primarily based on the momentum distributions of the knockout residue $^{30}$Mg. The parallel momentum distribution recorded in coincidence with $\gamma$ rays were then corrected for the feeding, assuming the present level scheme of Fig.~\ref{fig:levels}. In Fig.~\ref{fig:momdist}, the measured distributions are compared with reaction calculations based on the eikonal and sudden approximations. The theoretical approach adopted here is commonly used in studies of one-nucleon knockout reactions induced by a nuclear target and is well documented~\cite{TOS01,HAN03}. The parallel momentum distributions can be calculated from the residue- and neutron-target eikonal $S$ matrices, generated using double- and single-folding optical limit of Glauber's multiple scattering theory, respectively. As input to the calculation, densities of the residue and the target are required. Following the approach outlined in Refs.~\cite{GAD08,TOS14}, the density of $^{30}$Mg is taken from a Skyrme Hartree-Fock (HF) calculation using the SkX parameter set~\cite{BRO98}, while the density of the target nucleus $^9$Be is parameterized by a Gaussian with an RMS radius of \SI{2.36}{fm}. The removed-neutron radial wave functions, the one-nucleon overlaps, are calculated with the well-depth prescription where the parameters of a Woods-Saxon-plus-spin-orbit potential are constrained by the RMS radii of the orbitals from the spherical HF calculation. The model calculations are folded with the momentum distribution of the unreacted beam of $^{31}$Mg after passing through the target (\SI{140}{MeV/\clight} in FWHM), in order to account for the broadening due to the momentum profile of the incoming beam and straggling effects in the target. The theoretical momentum distributions are normalized to the measured counts in the momentum region \num{12.12}\text{--}\SI{12.56}{GeV/\clight}. The lower limit is chosen in order to avoid the influence of the low-momentum tail component, which cannot be described by the reaction model. In the following, spin-parity assignments to the observed states in $^{30}$Mg are discussed.

The states at \SI{1482}{keV} and \SI{1788}{keV} have been previously assigned as the $2_1^+$ and $0_2^+$ states, respectively. The measured momentum distributions, shown in Figs.~\ref{fig:momdist}(a) and (b), are compatible with knockout from the $1d_{3/2}$ and $2s_{1/2}$ orbitals, respectively. It should be noted that the present data did not allow for the clear extraction of the ground-state momentum distribution. This implies that there are remaining unobserved weak transitions that feed the ground state.

For the \SIhyp{2467}{keV} state, the present data clearly favor knockout from the $1d_{3/2}$ orbital [the reduced $\chi^2$ \num{4.9} is (\num{60.0}) for $1d_{3/2}$ ($2p_{3/2}$)], in contrast with the conclusion drawn in the previous study where a $2^-$ assignment was made~\cite{FER18}. This state was originally discussed as a candidate for the $2_2^+$ state~\cite{MAC05}, and in the present study, a $2^+$ assignment is firmly made.

For the state at \SI{3461}{keV}, the observed momentum distribution shows a narrow and almost symmetric shape. This is not compatible with the $4^+$ assignment previously given~\cite{DEA10}, as the spin coupling of the $1/2^+$ ground state in $^{31}$Mg with a neutron in the $sdpf$ orbitals does not allow a $4^+$ state to be populated directly. The momentum distribution suggests knockout from the $1d_{3/2}$ orbital, but the reduced $\chi^2$ [\num{14.2} (\num{43.1}) for $1d_{3/2}$ ($2p_{3/2}$)] is higher than in other cases. This may be explained by the presence of two near-degenerated states. A $\beta$-decay study proposed a doublet of states emitting \SI{1978}{keV} and \SI{1980}{keV} $\gamma$ rays~\cite{NIS16}, but in the present experiment it is not possible to disentangle the populations of these states because of the limited energy resolution. Nevertheless, a fit to the observed momentum distribution allows for the extraction of each contribution if assuming $1d_{3/2}$ and $2p_{3/2}$ components. The resulting cross sections are \SI{5.8(6)}{mb} and \SI{6.2(6)}{mb} for the $1d_{3/2}$ and $2p_{3/2}$ orbitals, respectively. One might assume a $2s_{1/2}$ component instead of $2p_{3/2}$, but such a state with a possible spin-parity of $1^+$ occurs only above \SI{4.8}{MeV} according to shell-model calculations (see Sec.~\ref{sec:sm}).

Because of the spin coupling of the $^{31}$Mg ground state and the $pf$ orbitals, only a single orbital contributes to the momentum distribution of a negative-parity state with $J^\pi = 1^-\text{--}4^-$. The removal of a neutron from the $1f_{7/2}$ orbital allows for the population of $3^-$ or $4^-$ states, and the \SIhyp{3302}{keV} momentum distribution supports knockout from this orbital. Given the observed sole branch to the $2_1^+$ state, a spin-parity assignment of $3^-$ with an E1 transition is most likely. The \SIhyp{4258}{keV} momentum distribution also undoubtedly shows knockout from the $1f_{7/2}$ orbital. This state decays via the \SIhyp{956}{keV} transition to the $3^-$ state. With the non-observation of decay branches to the $2^+$ states, this state likely has a spin-parity of $4^-$. The removal of a $2p_{3/2}$ neutron can lead to either $1^-$ or $2^-$ states. For the \SIhyp{3540}{keV} state, the momentum distribution is compatible with knockout from the $2p_{3/2}$ orbital, and a $1^-$ assignment is proposed here, because of the observation of the ground-state decay. This spin-parity assignment reasonably explains the branching ratio to the $2_1^+$ state ($\gamma$-ray intensity \SI{20(1)}{\percent} with respect to the ground-state transition). The newly-found \SIhyp{5210}{keV} state is also characterized by knockout from the $2p_{3/2}$ orbital, and guided by the absence of the ground-state decay, a spin-parity of $2^-$ is suggested.

The preferred population of high-spin states in fragmentation-like reactions~\cite{DEJ97,BEL00,YON01,CRA16} serves as a qualitative tool to identify such states. As can be seen in Fig.~\ref{fig:relxsec}, relative enhancements of the cross sections for the states at \SI{3379}{keV}, \SI{4183}{keV}, and \SI{4260}{keV} compared to the direct one-neutron knockout reaction from $^{31}$Mg are observed in the multi-nucleon removal reactions, suggesting that these states likely have high spins ($J\geq 4$). Furthermore, the corresponding momentum distributions in one-neutron knockout show shifted centroids (see Fig.~\ref{fig:momdist}(g) for the example of the \SIhyp{4260}{keV} state). This implies that these states are populated by non-direct one-neutron removal pathways~\cite{MUT16} such as the evaporation of one neutron following inelastic excitation. The \SIhyp{3379}{keV} state was assigned as a $4^+$ state in a previous study~\cite{DEA10}, and the present data add to the evidence for this assignment. It is worthwhile to emphasize that the $4_1^+$ state at \SI{3379}{keV} built on top of the $2_1^+$ state leads to $R_{42} = 2.3$, which is close to the vibrational limit. A candidate for an yrast state with $J = 5$~\cite{DEA10} was observed at \SI{4183}{keV} in the present measurement. Similarly, as can be seen in Fig.~\ref{fig:momdist}(g), the present data suggest a high-spin character of the \SIhyp{4260}{keV} state. The different momentum distributions for the \SIhyp{4258}{keV} and \SIhyp{4260}{keV} states prove that these are two different states.

\subsection{Spectroscopic factors}

Given a set of theoretical spectroscopic factors, the theoretical cross section for a final state $J_f^\pi$ is computed as the sum of contributions from each single-particle orbital with quantum numbers $nlj$
\begin{equation}
\sigma_\mathrm{th} = \sum_{nlj}
\left(\frac{A}{A-1}\right)^N C^2S(J_f^\pi,nlj)
\sigma_\mathrm{sp} (S_n+E_\mathrm{x}(J_f^\pi),nlj)
\end{equation}
where $\sigma_\mathrm{sp}$ is the single-particle cross section taken from the reaction calculation performed in the same framework as that described in the previous section. The $\sigma_\mathrm{sp}$ is calculated at the effective neutron separation energy $S_n+E_\mathrm{x}(J_f^\pi)$. The factor $[A/(A-1)]^N$, with $N$ being the major oscillator quantum number [2 (3) for the $sd$ ($pf$) orbitals], represents the center-of-mass correction to the shell-model spectroscopic factors. The experimental spectroscopic factors presented in Table~\ref{tab:sfactors} are calculated as the ratio of the measured cross section to the calculated $\sigma_\mathrm{sp}$. For $2^+$ states, the experimental spectroscopic factors are deduced assuming a pure $1d_{3/2}$ contribution. According to shell-model calculations detailed in Sec.~\ref{sec:sm}, the $1d_{3/2}$ component dominates over $1d_{5/2}$ for low-lying $2^+$ states.

The large deduced cross section populating the ground state leads to a sizable spectroscopic strength of \num{0.33(6)} by $2s_{1/2}$ neutron removal. This value should be seen as an upper limit, because of the presence of unplaced or unobserved transitions that feed the ground state. Such overestimation of exclusive cross sections for low-lying states is inevitable in $\gamma$-tagged measurements. A sizable spectroscopic factor of \num{0.39(5)} obtained for the $0_2^+$ state, is in line with a simple expectation that the $0_2^+$ state in $^{30}$Mg is characterized by substantial overlap with the intruder-dominated, deformed ground state of $^{31}$Mg. With the revised spin-parity assignment for the \SI{2467}{keV} state, the corresponding spectroscopic factor was determined to be \num{0.43(2)}. The existence of a doublet of states is proposed at \SI{3461}{keV}, and the spectroscopic factors for the two (unresolved) states are \num{0.32(3)} ($1d_{3/2}$) and \num{0.27(3)} ($2p_{3/2}$). The observed large contributions from the negative-parity orbitals are the direct consequence of the intruder-dominated character of the $^{30}$Mg ground state. This point is revisited in the next section.

\begin{table}[tb]
\caption{Experimental cross sections and spectroscopic factors. The orbital from which the neutron was removed is shown in the second column. The quoted uncertatinties include systematic contributions propagating from the inclusive cross section and uncertainties of $\gamma$-ray detection efficiency (assumed to be \SI{3}{\percent}).}
\label{tab:sfactors}
\begin{tabular}{S[table-format=4.0,table-align-text-post=false]cSSSSSS}
\toprule
{$E_\mathrm{x}$ (keV)} & {Orbital} & {$\sigma_\mathrm{exp}$ (mb)} & {$\sigma_\mathrm{sp}$ (mb)} & {$C^2S_\mathrm{exp}$} \\
\colrule
   0                     & $2s_{1/2}$ & 17.2(30){\footnotemark[1]} & 52.1 & 0.33(6){\footnotemark[1]} \\
1482                     & $1d_{3/2}$ & 10.2( 5)                   & 22.0 & 0.46(2) \\
1788                     & $2s_{1/2}$ & 13.8(19)                   & 35.0 & 0.39(5) \\
2467                     & $1d_{3/2}$ &  8.4( 4)                   & 19.8 & 0.43(2) \\
3302                     & $1f_{7/2}$ &  7.9( 4)                   & 17.1 & 0.46(2) \\
3379                     &            &  3.1( 2)                   &      &         \\
3461{\footnotemark[2]}   & $1d_{3/2}$ & 12.0( 5)                   & 18.0 & 0.67(3) \\
3461{\footnotemark[3]}   & $1d_{3/2}$ &  5.8( 6)                   & 18.0 & 0.32(3) \\
                         & $2p_{3/2}$ &  6.2( 6)                   & 22.7 & 0.27(3) \\
3540                     & $2p_{3/2}$ &  6.2( 3)                   & 22.5 & 0.27(1) \\
4183                     &            &  1.5( 1)                   &      &         \\
4258                     & $1f_{7/2}$ &  5.7( 3)                   & 16.1 & 0.35(2) \\
4260                     &            &  0.9( 1)                   &      &         \\
5210                     & $2p_{3/2}$ &  1.5( 1)                   & 18.9 & 0.08(1) \\
{Others\footnotemark[4]} &            &  8.4( 5)                   &      &         \\
\colrule
{Inclusive}              &            &  97 (3)                    &      &         \\
\botrule
\end{tabular}
\footnotetext[1]{These values should be seen as upper limits.}
\footnotetext[2]{A single state is assumed.}
\footnotetext[3]{A doublet is assumed.}
\footnotetext[4]{The sum of the states at \num{4683}, \num{4967}, \num{5022}, \num{5095}, \num{5414}, \num{5897}, and \SI{6064}{keV}. The placement in the level scheme is adopted from Ref.~\cite{SHI14}.}
\end{table}

\section{Discussion}

\subsection{Comparison to shell-model calculations}
\label{sec:sm}

\begin{figure}[tb]
\includegraphics[scale=0.5]{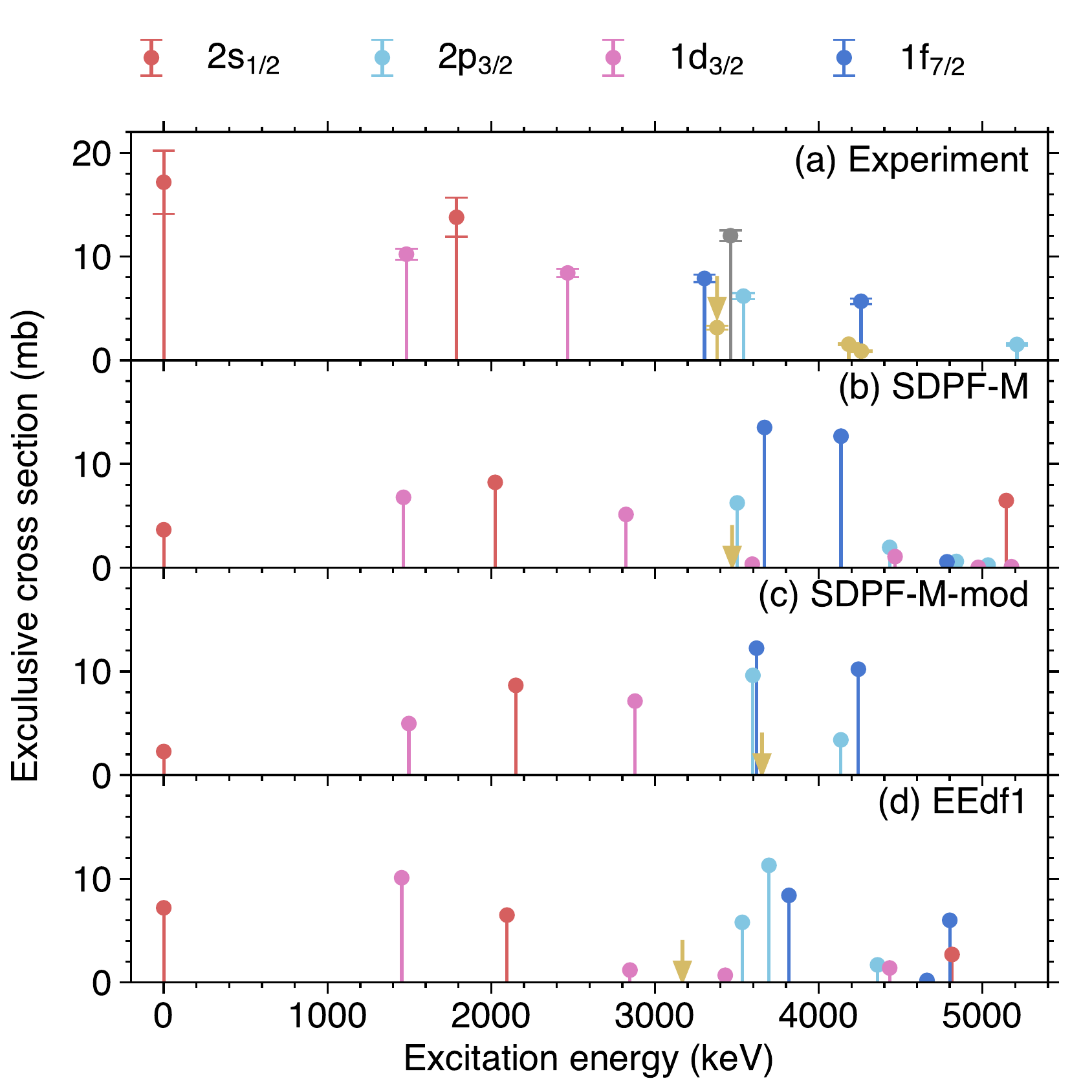}
\caption{Comparison of (a) experimental cross sections to those obtained from shell-model calculations using different interactions, (b) SDPF-M, (c) modified SDPF-M, and (d) EEdf1. States characterized by $2s_{1/2}$, $2p_{3/2}$, $1d_{3/2}$, and $1f_{7/2}$ are respectively shown in red, cyan, pink, and blue. The experimental ground-state cross section should be seen as an upper limit. Two close-lying states that cannot be resolved in the present experiment likely contribute for the experimental cross section of the \SIhyp{3461}{keV} state shown in gray (see text for details). The yellow data points in the panel (a) are for the observed candidates for the high-spin states. The yrast $4^+$ states are indicated by the yellow arrows.}
\label{fig:xsec}
\end{figure}

To assess the theoretical description of $^{30}$Mg, large-scale shell-model calculations were performed using the SDPF-M~\cite{UTS99} and EEdf1~\cite{TSU17} effective interactions. The former interaction, developed in 1999, has been traditionally used in studies of island-of-inversion nuclei. The model space of the interaction comprises of the full $sd$ shell and the lower half of the $fp$ shell ($1f_{7/2}$ and $2p_{3/2}$) for both neutrons and protons. The two-body matrix elements (TBMEs) and the single-particle energies (SPEs) were empirically adjusted to reproduce selected experimental observables. The EEdf1 interaction was developed most recently with the model space extending to the full $fp$ shell. This interaction is different from SDPF-M by construction in that the TBMEs are microscopically derived. The Kuo-Krenciglowa method and the Entem-Machleidt QCD-based nucleon-nucleon interaction are used for the derivation of TBMEs, while the SPEs are fitted to reproduce selected experimental data. The calculations were performed with the code \textsc{kshell}~\cite{SHI13}. The SDPF-M result is obtained from the full diagonalization, whereas the number of excitations is restricted up to 6p6h for the EEdf1 interaction, due to computational limitations. The calculated levels and their cross sections are compared with those measured in Fig.~\ref{fig:xsec}. The overall level structure, especially the locations of the $2_1^+$, $0_2^+$, and $2_2^+$ states, as well as the negative-parity levels, are remarkably well reproduced by both calculations. The predicted excitation energies of the $4_1^+$ state using the SDPF-M and EEdf1 interactions are \SI{3.47}{MeV} and \SI{3.16}{MeV}, respectively, showing good agreement with the experiment (\SI{3379}{keV}). We note that the recently-developed SDPF-U-MIX interaction~\cite{CAU14}, also well reproduces the experimental level scheme (see Ref.~\cite{FER18}).

In addition to the excitation energies, the calculated electromagnetic transition strengths between low-lying states show good agreements with the experimental data. As shown in Table~\ref{tab:elemag}, the theoretical and experimental strengths compare well for the $2_1^+\to0_1^+$ and $0_2^+\to2_1^+$ transitions. Experimentally, a substantial M1 component of the $2_2^+\to2_1^+$ transition was suggested, following the upper limit on the $2_2^+$ level half-life of \SI{5}{ps}~\cite{MAC05}. The calculations indicate that the $2_2^+\to2_1^+$ transition has strong mixing of M1. Simulations show that a half-life of \SI{5}{ps} leads to a slight shift of the \SIhyp{985}{keV} peak by \SI{1}{keV} in the Doppler-corrected spectrum produced with the mid-target velocity. The observation of the peak at \SI{985}{keV} is, therefore, in line with the above-mentioned upper limit.

\begin{table}[tb]
\centering
\caption{Comparison of theoretical and experimental transition strengths. Adopted effective charges are $(e_p,e_n) = (1.3,0.5)e$ for SDPF-M~\cite{UTS99} and $(e_p,e_n) = (1.25,0.25)e$ for EEdf1~\cite{TSU17}. The strengths are taken to be in units of \si{\elementarycharge^2.fm^4} (E2) and $\mu_N^2$ (M1).}
\label{tab:elemag}
\begin{tabular}{cccc}
\toprule
Transition & SDPF-M & EEdf1 & Experiment \\
\colrule
$0_1^+\to2_1^+$ & 332       & 285       & \num{241(31)}\footnotemark[1] \\
$0_2^+\to2_1^+$ & 38        & 51        & \num{53(6)}\footnotemark[2]   \\
$2_2^+\to2_1^+$ & 1.8 (E2)  & 1.0 (E2)  & \num{>123}\footnotemark[3]    \\
                & 0.16 (M1) & 0.19 (M1) &  \\
\botrule
\end{tabular}
\footnotetext[1]{Reference~\cite{NIE05}.}
\footnotetext[2]{Reference~\cite{SCH09}.}
\footnotetext[3]{The lower limit is given by the upper limit on the lifetime when a pure E2 character is assumed~\cite{MAC05}.}
\end{table}

Unlike the level structure and the transition strengths, exclusive cross sections indicated by bars in Fig.~\ref{fig:xsec} show marked differences depending on the interaction. As the SDPF-M interaction does not reproduce the correct level ordering of $^{31}$Mg~\cite{TSU17}, the wave function of the lowest $1/2^+$ state is taken as the $^{31}$Mg ground state for the calculation of the spectroscopic factors. We note that the EEdf1 interaction correctly reproduces the ground-state spin-parity, and the calculation and the experimental level scheme of $^{31}$Mg agree well~\cite{NIS19}. As can be seen in Fig.~\ref{fig:xsec}, the comparison with the measurement does not allow for a definitive choice between the shell-model calculations; the spectroscopic factors associated with the negative-parity states are quantitatively better reproduced in the EEdf1 calculation, whereas the large spectroscopic factor for the $2_2^+$ state is only reproduced by the SDPF-M interaction.

We note that the theoretical inclusive cross section calculated with the SDPF-M interaction amounts to \SI{91}{mb} with a theoretical uncertainty of \SI{13}{mb} estimated by varying the $S_n$ by $\pm\SI{500}{keV}$. When using the EEdf1 interaction, the number of states calculated is limited due to the high computational cost and therefore the inclusive cross section could not be computed. The experimental inclusive cross section for direct knockout is estimated to be \SI{91(3)}{mb} by subtracting the cross sections of the \SIhyp{3379}{keV}, \SIhyp{4183}{keV}, and \SIhyp{4260}{keV} states, populated by non-direct one-neutron removal, from the measured inclusive cross section. According to the systematics of Ref.~\cite{TOS14}, the ratio of the experimental to the theoretical direct-knockout inclusive cross sections is expected to be around \num{0.9}. The present result conforms to this expectation within the uncertainty margin.

In order to see the correlation between the interaction and the predicted spectroscopic factors, calculations with modified SPEs have been performed based on the SDPF-M interaction~\cite{HIM08,XU18,KAN10}. The overestimated $1f_{7/2}$ spectroscopic factor is mitigated by reducing the energy spacing between the $1f_{7/2}$ and $2p_{3/2}$ orbitals corresponding to the $N=28$ gap. This is demonstrated in Fig.~\ref{fig:xsec}, where ``SDPF-M-mod'' represents the calculation with the single-particle energies of the $1f_{7/2}$ ($2p_{3/2}$) orbitals raised (lowered) by \SI{0.5}{MeV}. This modification reduces the $1f_{7/2}$ neutron occupancy in the $^{31}$Mg ground state by \SI{14}{\percent}, leading to smaller $1f_{7/2}$ spectroscopic factors. This indicates that the spectroscopic strengths of the negative-parity states are also sensitive to the effective shell-gap size. Note that a lowering of the $2p_{3/2}$ SPE alone~\cite{HIM08,XU18,KAN10} also leads to a reduced $1f_{7/2}$ occupancy, but results in a smaller $2_2^+$ spectroscopic strength.

\begin{figure}[tb]
\includegraphics[scale=0.5]{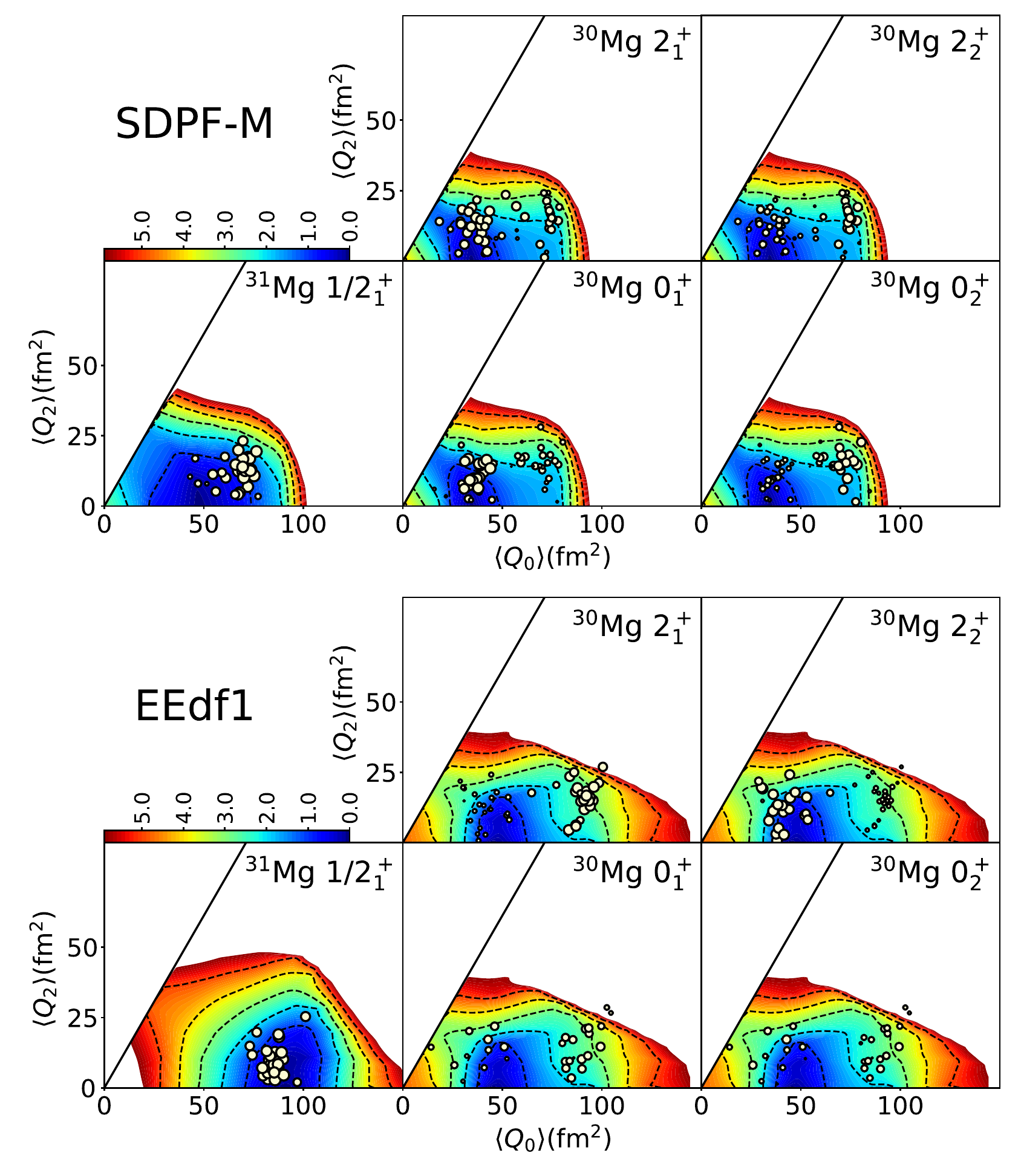}
\caption{T-plots for the $0_1^+$, $2_1^+$, $0_2^+$, and $2_2^+$ states in $^{30}$Mg and the lowest $1/2^+$ state in $^{31}$Mg produced using the SDPF-M and EEdf1 interactions. The contour plots show the potential energy surfaces, while the circles indicate the deformed Slater determinants (see text for details).}
\label{fig:tplots}
\end{figure}

To further differentiate the two interactions used in the present work, the low-lying states in $^{30}$Mg, i.e.\ the $0_1^+$, $2_1^+$, $0_2^+$, and $2_2^+$ states, were analyzed by the T-plot technique~\cite{TSU14} to explore the nature of the underlying wave functions. The plots, in Fig.~\ref{fig:tplots}, show the potential energy surface (PES) obtained by a constrained Hartree-Fock calculation using the shell-model interactions. The circles in the plots represent deformed Slater determinants in the Monte Carlo Shell Model (MCSM) framework~\cite{OTS01}, and their size is proportional to the overlap between the MCSM eigenstate and the deformed Slater determinant. With this approach, the intrinsic deformation of each state is visualized. As can be seen in Fig.~\ref{fig:tplots}, the two interactions predict very different structures for $^{30}$Mg despite the similar level schemes, but the shape-coexisting feature persists; the circles are localized both in the weakly-deformed and strongly-prolate-deformed sides. For the SDPF-M calculation, the $0_1^+$ state exhibits a weakly-deformed shape, indicated by the clustering of large circles in the region of $\langle Q_0\rangle\approx\SI{30}{fm^2}$ and $\langle Q_2\rangle\approx\SI{10}{fm^2}$, while the $0_2^+$ is governed by strongly deformed configurations where the circles are localized around $\langle Q_0\rangle\approx\SI{70}{fm^2}$. The $2_1^+$ and $2_2^+$ states are characterized by a higher degree of shape mixing. Here, it is worth noting that the calculated spectroscopic factors may be intuitively understood by the T-plots. The ground state of $^{31}$Mg is prolate-deformed, and the $0_1^+$ ($0_2^+$) states are characterized by weakly- (strongly-) deformed configurations, thus yielding small (large) overlaps and spectroscopic factors. The spectroscopic factors of the $2_1^+$ and $2_2^+$ states with similar magnitudes are understood by large shape fluctuations. In the EEdf1 results, it is seen that shape mixing is more pronounced in the $0_1^+$ and $0_2^+$ states. Moreover, the shape coexistence in the $2_1^+$ and $2_2^+$ states is ``inverted'' with respect to the na\"ive expectation, i.e.\ the $2_1^+$ state is dominated by deformed configurations while the $2_2^+$ state is close to spherical.

Usually, excitation energies are used as a first test of theoretical calculations. The present results demonstrate that more detailed experimental information, such as spectroscopic factors, is required to further differentiate the models. The findings obtained here provide guidance for refinements of the interactions, and the new experimental information will serve as a benchmark towards the full description of the nuclear structure in and around the island of inversion.

\subsection{Transition into the island of inversion}

\begin{figure}[tb]
\includegraphics[scale=0.5]{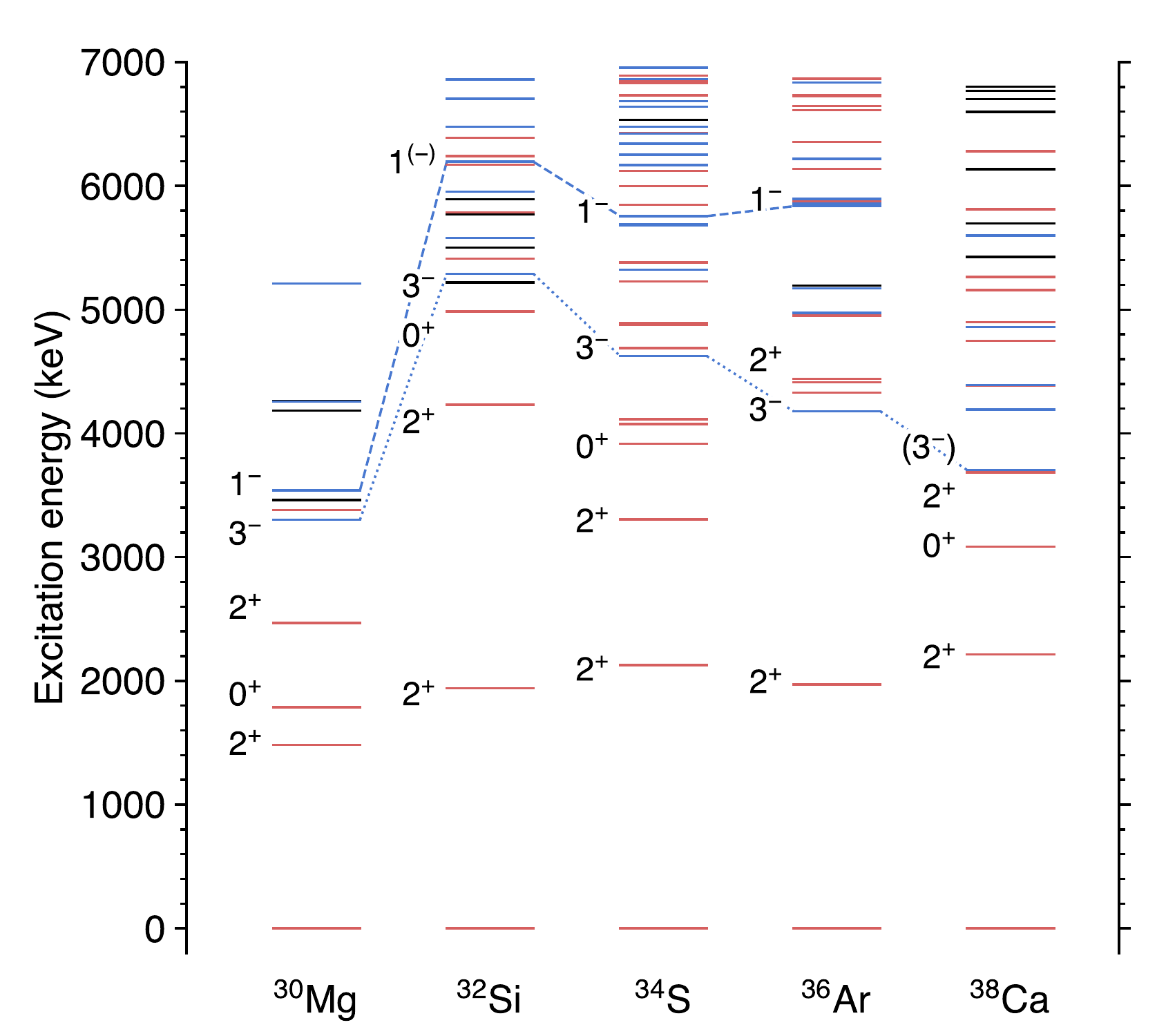}
\caption{Systematics of levels in $N=18$ isotones. Level data are taken from the ENSDF database, while the levels in $^{30}$Mg are taken from the present analysis. The lowest $1^-$ ($3^-$) states are connected by bashed (dotted) lines. Negative-parity (positive-parity) states are shown in blue (red).}
\label{fig:systematics}
\end{figure}

Systematics of the level structure along the $N=18$ isotones are displayed in Fig.~\ref{fig:systematics}. It is seen that, going from $^{38}$Ca to $^{32}$Si, the $1^-$ states stay constant around \SI{6}{MeV} and the $3^-$ states continue to rise steadily up to \SI{5.3}{MeV}, but a rapid lowering of the excitation energies is observed at $^{30}$Mg. This is interpreted as a precursory structural change approaching the island of inversion. If one assumes a na\"ive 1p1h excitation in the framework of the shell model, the excitation energy of the negative-parity state can be related to the gap size between effective single-particle orbitals. A $1^-$ state is formed by promoting one neutron from the $1d_{3/2}$ orbital to $2p_{3/2}$. Therefore, the observed drop of the $1^-$ states could be interpreted as driven by the reduced gap between the $1d_{3/2}$ and $2p_{3/2}$ orbitals, thus indicating the shell evolution that is considered to be responsible for the appearance of the island of inversion. This behavior is found to coincide with the lowering of the effective single-particle energy of the $2p_{3/2}$ orbital with respect to $1d_{3/2}$ both in the SDPF-M and EEdf1 interactions. A drop of excitation energy is also seen for the $3^-$ states, but the underlying configuration is not as simple as the $1^-$ states, as a $3^-$ state can be by promoting one neutron from the $1d_{3/2}$ orbital to both the $1f_{7/2}$ or $2p_{3/2}$ orbitals. This drop in excitation energy is analogous to the $N=8$ chain where the sudden lowering of the $1^-$ state is observed at $^{12}$Be~\cite{IWA00}.

\begin{figure}[tb]
\includegraphics[scale=0.5]{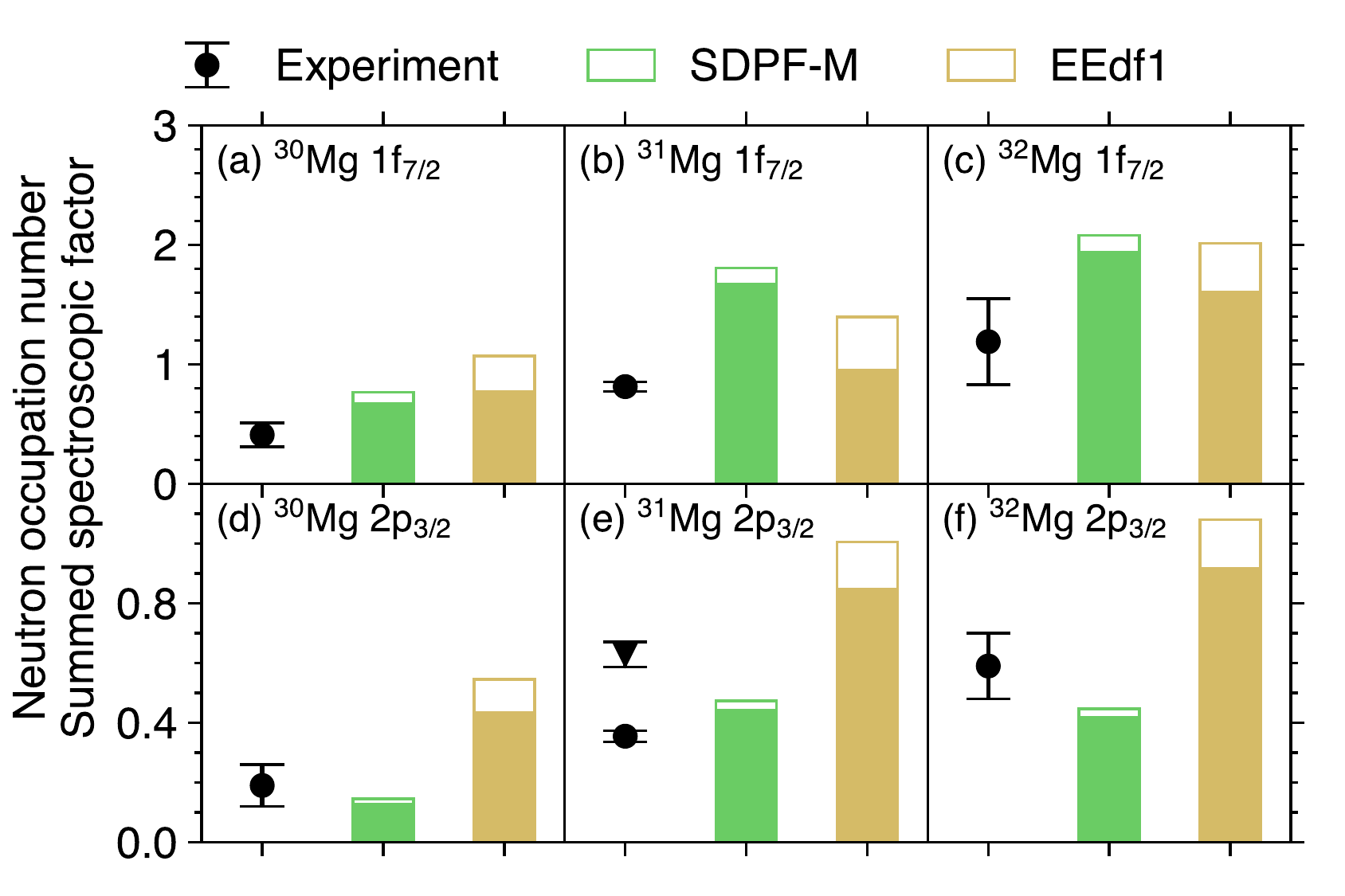}
\caption{Comparison of summed experimental spectroscopic factors to neutron occupancies calculated for (a) the $1f_{7/2}$ and (b) the $2p_{3/2}$ orbitals. The filled region shows calculated spectroscopic factors summed up to $S_n$. The calculated values are corrected for the $[A/(A-1)]^N$ factor. The triangle represents the sum including the contribution of $2p_{3/2}$ at \SI{3461}{keV} where a doublet is proposed.}
\label{fig:occupation}
\end{figure}

To track the evolution of the $fp$-shell occupancy approaching the island of inversion, the summed spectroscopic factors associated with the $1f_{7/2}$ and $2p_{3/2}$ orbitals are shown in Fig.~\ref{fig:occupation}. For $^{31}$Mg, the spectroscopic factors populating negative-parity states in $^{30}$Mg were taken from the present work. Those of $^{30}$Mg and $^{32}$Mg were taken from a previous measurement of one-neutron knockout reactions~\cite{TER08}. The observed continuous transition expands the early concept of the island of inversion with sharp borders~\cite{WAR90} and corroborates, in a quantitative manner, a gradual-transition scenario which has been theoretically predicted~\cite{UTS99}.

\section{Summary and conclusion}

The structure of the neutron-rich nucleus $^{30}$Mg located at the boundary of the island of inversion was studied in detail by in-beam $\gamma$-ray spectroscopy. To populate states in $^{30}$Mg, the one-neutron knockout reaction from $^{31}$Mg was primarily employed. Multi-nucleon removal reactions leading to $^{30}$Mg were also studied to obtain additional information beneficial to construct an updated level scheme and to constrain the spins and parities of populated states. In the knockout reaction, the momentum distributions of the residual $^{30}$Mg are characteristic of the angular momentum of the removed neutron. Together with the well-established reaction mechanism and its theoretical prescriptions, the momentum distributions were then used to assign spins and parities of the populated states. Spectroscopic factors for each state were also deduced based on the measured cross sections and reaction model calculations.

The location of the negative-parity states, which reflects the effective size of the $N=20$ shell gap, was established. Contrary to previous studies, indications of negative-parity states at low excitation energies below \SI{3}{MeV} were not found. Nevertheless, the drop of the excitation energy of the negative-parity states from Si ($Z = 14$) to Mg ($Z = 12$) pertains. The negative-parity states were strongly populated in the knockout reaction. In particular, the neutron knockout from the $2p_{3/2}$ orbital showed significant spectroscopic strength, corroborating the intruder-dominated configuration in the ground state of $^{31}$Mg. These experimental findings are
interpreted as a precursory structural change approaching the island of inversion.

To gain more insight into the nuclear structure of $^{30}$Mg, large-scale shell-model calculations have been performed. The updated level scheme is very well reproduced by the calculations, whereas the spectroscopic factors show large variations depending on the interaction used. It was also found that the picture of shape coexistence in $^{30}$Mg largely depends on the interaction and is correlated with the calculated spectroscopic factors. The observed spectroscopic factors are not completely reproduced by the shell-model calculations using the SDPF-M and EEdf1 interactions for all states, implying the transition into the island of inversion is not fully captured in the present shell model.

To fully map the transition into the island of inversion, further detailed spectroscopic studies of nuclei in and around the island of inversion are needed. Transfer reactions starting from or leading to $^{30}$Mg could be used to verify the predicted small spectroscopic factors between the two ground states, as the reactions are performed using the missing-mass method. Furthermore, muti-step Coulomb excitation could shed more light on the shape coexistence and deformation in $^{30}$Mg.

\begin{acknowledgments}
We would like to express our gratitude to the accelerator staff at NSCL for their efforts in beam delivery during the experiment. N.~K.\ wish to acknowledge support of the Grant-in-Aid for JSPS Fellows (18J12542) from the Ministry of Education, Culture, Sports, Science, and Technology (MEXT), Japan. The shell-model calculations were enabled by the CNS-RIKEN joint project for large-scale nuclear structure calculations and were performed mainly on the Oakforest-PACS supercomputer. N.~S.\ acknowledges support from ``Priority Issue on post-K computer'' (hp190160) and ``Program for Promoting Researches on the Supercomputer Fugaku'' (hp200130) by JICFuS and MEXT, Japan. J.~A.~T.\ acknowledges support from the U.K.\ Science and Technology Facilities Council Grant No.\ ST/L005743/1. This work was supported by the U.S.\ Department of Energy (DOE), Office of Science, Office of Nuclear Physics, under Grant No.\ DE-SC0020451 and by the U.S.\ National Science Foundation (NSF) under Grant No.\ PHY-1306297. GRETINA was funded by the U.S.\ DOE, Office of Science. Operation of the array at NSCL is supported by the U.S.\ NSF under Cooperative Agreement No.\ PHY-1102511 (NSCL) and DOE under Grant No.\ DE-AC02-05CH11231 (LBNL).
\end{acknowledgments}

%\bibliography{mg30prc.bib}

%

\end{document}